\documentclass[aps,prl,twocolumn,groupedaddress]{revtex4-1}

\usepackage{graphicx}

\begin{document}

\title{Elasticity-driven collective motion in active solids and active crystals}

\author{Eliseo Ferrante}
\affiliation{Laboratory of Socioecology and Social Evolution, Katholieke Universiteit Leuven, Leuven, Belgium}
\author{Ali Emre Turgut}
\affiliation{University of Turkish Aeronautical Association, Ankara, Turkey}
\author{Marco Dorigo}
\affiliation{IRIDIA, Universit\'e Libre de Bruxelles, Brussels, Belgium}
\author{Cristi\'an Huepe}
\affiliation{614 N. Paulina St., Chicago IL 60622-6062, USA}
\date{\today}

\begin{abstract}
We introduce a simple model of self-propelled agents connected by linear springs, with no explicit alignment rules. Below a critical noise level, the agents self-organize into a collectively translating or rotating group. We derive analytical stability conditions for the translating state in an elastic sheet approximation. We propose an elasticity-based mechanism that drives convergence to collective motion by cascading self-propulsion energy towards lower-energy modes. Given its simplicity and ubiquity, such mechanism could play a relevant role in various biological and robotic swarms.
\end{abstract}


\maketitle



Animal groups that move together, such as bacterial colonies, insect swarms, bird flocks, or fish schools \cite{Breder54,Okubo86,Grunbaum94,ShapiroBook97,Couzin2002,SumpterBook}, 
are all examples of biological systems displaying Collective Motion (CM).
In recent years, the dynamics of such systems (referred to here generically as {\it swarms}) has been the subject of intense research
\cite{SumpterBook,DeThViReview2012,VicsekZafeirisReview2012,SwarmRoboticsReview2013}. 
A number of theoretical models have been introduced to study swarms and to develop control rules that achieve similar coordinated collective dynamics in groups of autonomous robots
\cite{VicsekZafeirisReview2012,SwarmRoboticsReview2013}.
Despite this proliferation of CM algorithms, there is still no comprehensive understanding of the underlying mechanisms that can lead a group of self-propelled agents to self-organize and move in a common direction.

The current CM paradigm has been strongly influenced by the seminal work of Vicsek et al.~\cite{ViCzPRL95}, which introduced a minimal model for flocking, the {\it Vicsek model}, that has become a referent in the field 
\cite{VicsekZafeirisReview2012,DeThViReview2012,SwarmRoboticsReview2013}.
This model describes a group of point particles advancing at a fixed common speed, only coupled through alignment interactions that steer them towards the mean heading direction of all particles within a given radius
\cite{ViCzPRL95,CzViJPhysA1997,ChRaPRE2008}.
In this framework, a swarm can be viewed as a group of self-propelled spins with aligning interactions, described by a variation of the XY-model \cite{CriticalPhenomenaBook} where spins advance in their pointing direction rather than remaining affixed to a lattice.
In the continuous limit, this system becomes a fluid of self-propelled spins that follows the hydrodynamic theory developed in \cite{ToTuPRL95,ToTuPRE98,PhysRevE.74.022101,PhysRevLett.109.098101}.
More recently, other models that do not rely on explicit alignment interactions have been introduced. 
In \cite{RoCoPRL09}, for example, CM is driven by escape-pursuit interactions only; in \cite{GrArNJP08}, by inelastic collisions between isotropic agents; and in \cite{SzSzPRE06} and \cite{HeFiPRE11}, by short-range radial forces coupled to each agent's turning dynamics.
Given that Vicsek-like algorithms rely on explicit alignment rules to achieve CM \cite{ViCzPRL95,Couzin2002,Gregoire2003157,GrChPRL2004}, it was initially surprising that such systems could self-organize without them.
While it can be argued that all these models include at least an {\it implicit} alignment interaction, it remains unclear if they are all driven to CM by the same underlying mechanisms and to what extent agents must exchange orientation information, either explicitly or implicitly, to achieve CM.

In this Letter, we introduce a CM mechanism that is based on a very different paradigm: the emergence and growth of regions of coherent motion due to standard elasticity processes.
We explore this mechanism by introducing a simple two-dimensional Active Elastic Sheet (AES) model with spring-like interactions between neighboring agents and no explicit alignment, which describes what we refer to as an {\it active solid} or an {\it active crystal}.


We define the AES model as a system of $N$ agents on a two-dimensional plane, where the position $\vec{x}_i$ and orientation $\theta_i$ of each agent $i$ follow the overdamped equations of motion
\begin{eqnarray}
\label{Eq:dynX}
\dot{\vec{x}}_i &=& v_0 \, \hat{n}_i + 
			\alpha \, \left[ \left( \vec{F}_i +  D_r \, \hat{\xi}_r \right) \cdot \hat{n}_i  \right] \, 
													\hat{n}_i \mbox{,} \\
\label{Eq:dynTheta}
\dot{\theta}_i &=& \beta \, \left[ \left( \vec{F}_i +  D_r \, \hat{\xi}_r \right) \cdot \hat{n}_{i}^{\perp} \right] + 
											D_{\theta} \, \xi_{\theta} \mbox{.}
\end{eqnarray}
Here, $v_0$ is the forward biasing speed that induces self-propulsion (injecting energy at the individual particle level), $\hat{n}_i$ and $\hat{n}_{i}^{\perp}$ are two unit vectors pointing parallel and perpendicular to the heading direction of agent $i$, and parameters $\alpha$ and $\beta$ are the inverse translational and rotational damping coefficients, respectively.
The total force over agent $i$ is given by
$ \vec{F}_i = \sum_{j \in S_i} \left( - k / l_{ij} \right) \left( \left| \vec{x}_i - \vec{x}_j \right| - l_{ij} \right) $,
a sum of linear spring-like forces with equilibrium distances $l_{ij}$ and spring constants $k/l_{ij}$.
Each set $S_i$ contains all agents interacting with agent $i$ and remains fixed throughout the integration. 
This system is thus akin to a spring-mass model of elastic sheet \cite{fetter2003theoretical} where masses are replaced by self-propelled agents that turn according to $\vec{F}_i \cdot \hat{n}_{i}^{\perp}$ and move forward or backwards following $\vec{F}_i \cdot \hat{n}_i$ and their self-propulsion.
We include {\it actuation noise} (fluctuations of the individual motion) by adding $D_{\theta} \, \xi_{\theta}$ to the heading angle, where $D_{\theta}$ is the noise strength coefficient and $\xi_{\theta}$ a random variable with standard, zero-centered normal probability distribution of variance $1$. 
We include {\it sensing noise} (errors in the measured forces) by adding $D_r \, \hat{\xi}_r$ to $\vec{F}_i$, with $D_r$ the noise strength coefficient and $\hat{\xi}_r$ a randomly oriented unit vector.
The degree of alignment in the system is monitored through the usual polarization order parameter 
\begin{equation}
\label{Eq:psi}
\psi = \frac{1}{N} \left\| \sum_{i = 1}^N \hat{n_i} \right\|.
\end{equation}
If all agents are perfectly aligned, we have $\psi = 1$. If agents are instead randomly oriented or rotating about the group's barycenter, we have $\psi = 0$.

The AES model was designed to study CM under conditions that are in many ways opposite to those in the Vicsek model.
While the only information that agents exchange there is their relative heading angle, here they only sense their relative positions. 
While changing interacting neighbors over time has been shown to be necessary there for achieving long-range order 
\cite{CzViJPhysA1997,ToTuPRL95,AlHuJSP2003}, here virtual springs connect the same agents throughout the dynamics.
Furthermore, while both models describe overdamped systems, these are implemented differently. 
In the Vicsek model, particles switch instantaneously to the desired heading angle; in the AES case, we integrate instead the overdamped angular equation (\ref{Eq:dynTheta}), which turns out to be necessary here for achieving CM.

We integrate Eqs.~(\ref{Eq:dynX}) and (\ref{Eq:dynTheta}) numerically using a standard Euler method.
All simulations below are carried out with $\alpha = 0.01$, $\beta = 0.12$, $v_0 = 0.002$, and $dt = 0.1$.

%
\begin{figure}
\includegraphics[width=86mm]{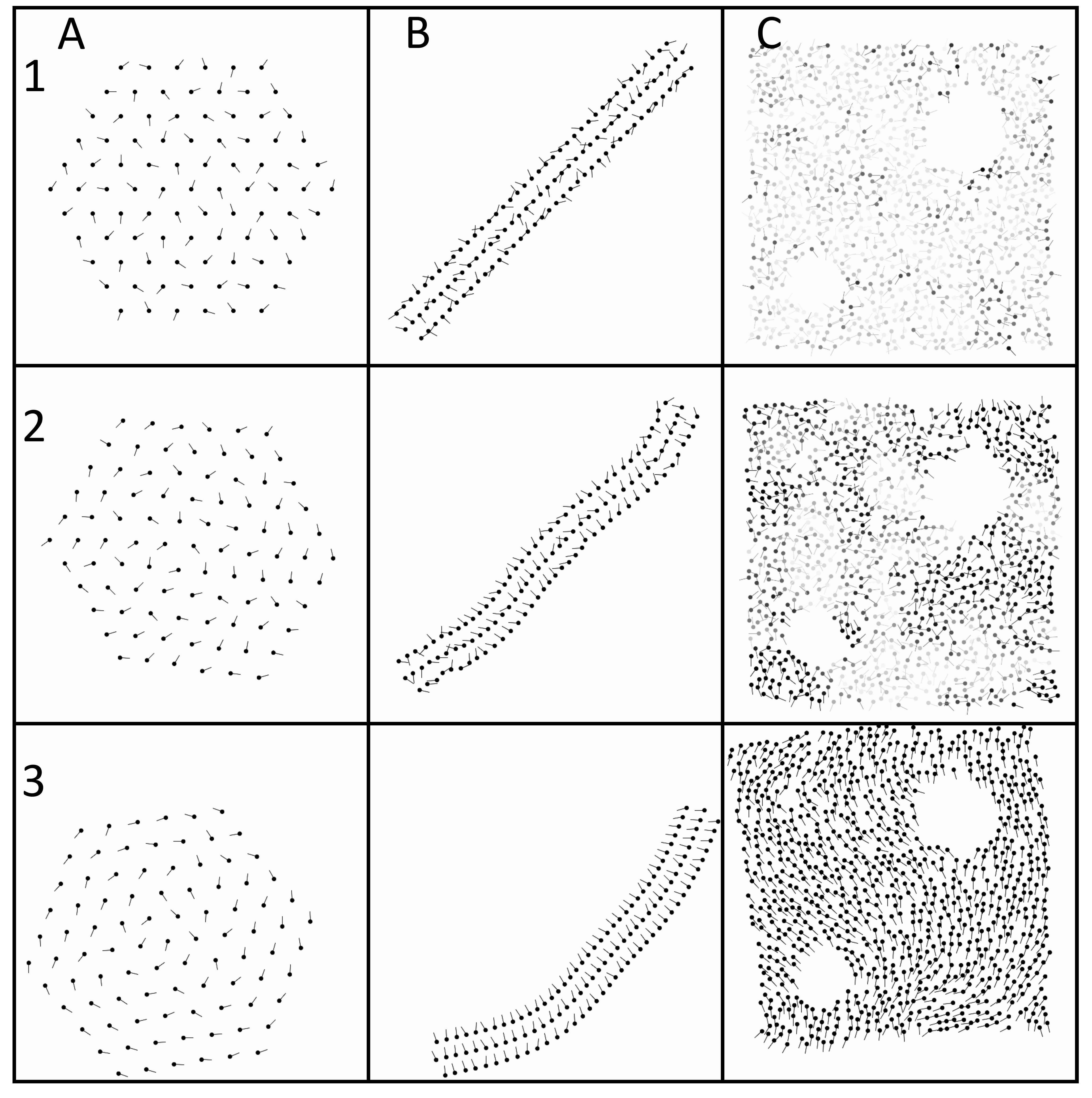}
\caption{\label{Fig1}
Active elastic sheet simulations of Eqs.~(\ref{Eq:dynX}) and (\ref{Eq:dynTheta}). 
A: Hexagonal active crystal at $t=0$ (A1), $240$ (A2), and $1700$ (A3).
B: Rod-like active crystal at $t=0$ (B1), $400$ (B2), and $1700$ (B3).
C: Active solid at same times as column B; darker agents symbolize higher local alignment.
}
\end{figure}
%

Figure 1 presents three runs of the AES model.
Column A displays the dynamics of an hexagonal active crystal composed of $N=91$ agents.
At $t=0$ (panel A1), agents are placed with random orientations on a perfect hexagonal lattice, separated by 
$d_A = 0.65$.
Nearest neighbors are connected by springs with natural length $l = d_A$ and spring constant $k/l=5/0.65$.
Only sensing noise is considered here ($D_r = 0.5$, $D_{\theta} = 0$), but results remain qualitatively unchanged for other types of noise.
As time advances, growing regions of coherent motion develop, eventually deforming the whole structure (A2) until the group starts translating or rotating collectively. 
In the case shown, the system converges to a rotating state where the axis of rotation and barycenter do not coincide (A3), thus rotating while translating.
Note that rotating states will always have higher elastic energy, since inner and outer shells cannot move at the same $v_0$ speed and must be sped up or slowed down by elastic forces.
They are less frequent and metastable, eventually relaxing to lower-energy, translating solutions.
However, we show one here to illustrate its dynamics, which cannot be attained by the Vicsek model.

Column B displays an active elastic rod, comprised of $N=118$ agents arranged into three rows, for the same noise as in column A. 
It is generated (B1) by placing randomly oriented agents with nearest-neighbor distances $d_B = 0.32$ (within rows) and $d_B^* = 0.58$ (between rows), linking all agents separated by $d<1$ with springs of natural length $d$ and spring constant $k/l=5/d$.
Here again, larger and larger regions of coherent deformation emerge (B2) until CM is attained and the rod starts moving (B3). Since the first bending mode has the largest final deformation, a collective heading direction perpendicular to the rod's axis is favored.

Column C shows $N=891$ agents forming an active solid (given the irregular agent positions) with two holes.
To construct it, we distribute agents at random, homogeneously within the structure, connecting all agents separated by $d<1$ with springs, following the same procedure used in column B.
Noise is set here to zero, but equivalent dynamics are observed for small enough $D_r$ and $D_{\theta}$.
To highlight ordered regions, each agent's darkness is displayed proportional to the local order, defined similar to $\psi$ but summing only over the focal agent and others linked to it, instead of the whole system.
Initially, most of the structure appears in light gray, since agents are randomly oriented (C1). 
As time advances, growing regions of coherent motion emerge (C2), until the whole structure starts moving when agents become sufficiently aligned (C3).

%
\begin{figure}[t]
\includegraphics[width=86mm]{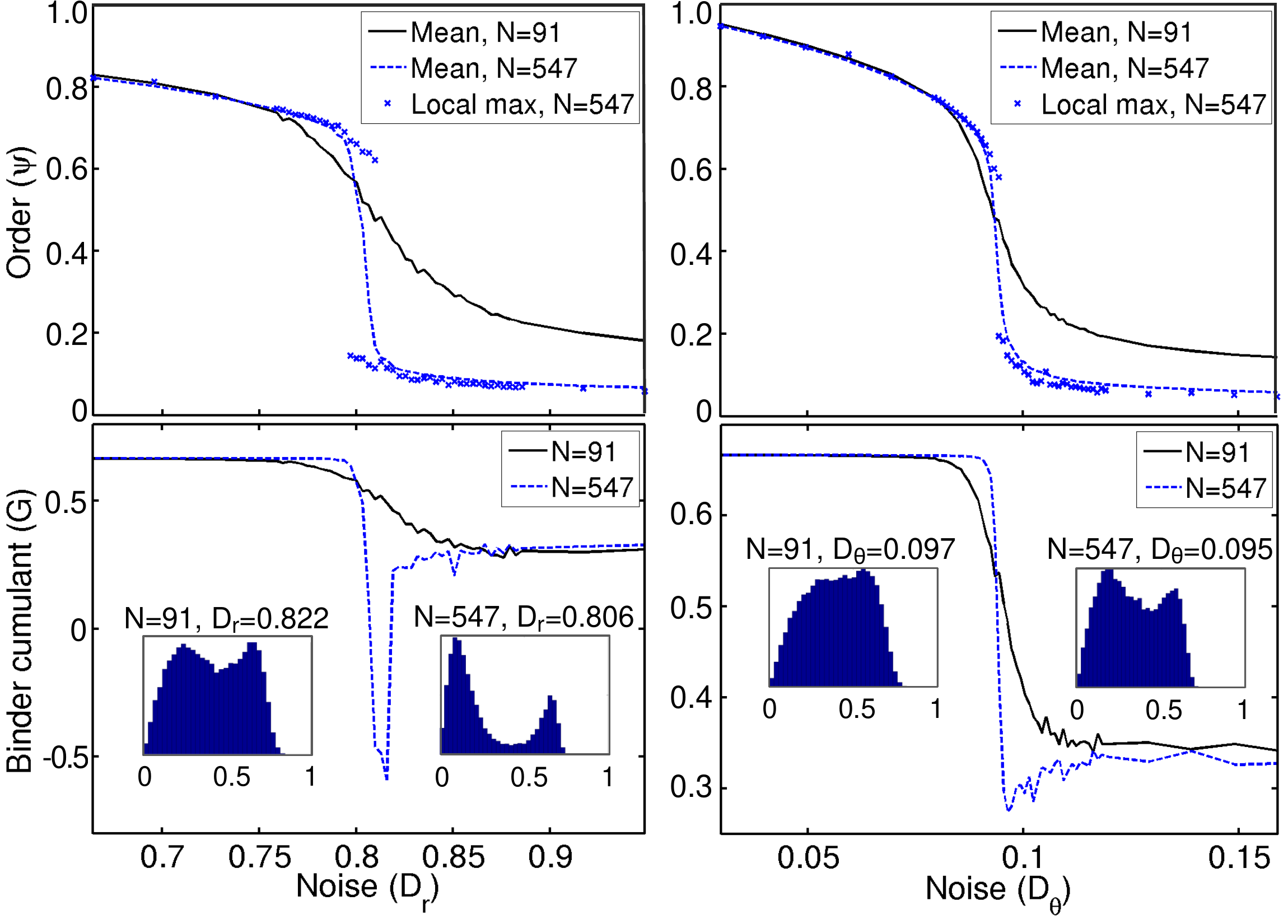}
\caption{\label{Fig2} (color online).
Order parameter and Binder cumulant vs. positional sensory noise $D_r$ and angular actuation noise $D_\theta$ 
for hexagonal active crystals with $N=91$ (same as on Fig.~\ref{Fig1}, panel A1) and $N=547$ agents. 
Top panels display the mean and local maxima of the distribution of $\psi$ values obtained in simulations. 
Insets show these distributions in the transition region.
For large enough systems, both cases display a first order transition with a bistable region.
}
\end{figure}
%

The AES model displays a discontinuous order-disorder transition similar to that in the Vicsek model.
Figure \ref{Fig2} examines this transition as a function of noise for the hexagonal active crystal on column A of Fig.~\ref{Fig1} and for a larger ($N=547$) hexagonal configuration with identical parameters.
We performed $30$ to $80$ runs per noise value, storing $2000$ values of $\psi$ per run (every $500$ time-steps after the initial $10^6$).
Top panels show the mean and local maxima of the $\psi$ distributions and bottom ones, the corresponding Binder cumulants 
$G = 1 - \frac{1}{3} \left< \psi^4 \right> / \left< \psi^2 \right>^{2} $ \cite{BiRPP1997}.
As we increase either sensing noise $D_r$ (left) or actuation noise $D_{\theta}$ (right), $\psi$ jumps from an ordered state where agents self-organize to a disordered state where they continue to point in random directions.
The displayed Binder cumulants and bimodal distributions show that there is a region of bistability around the critical point in both cases.
In the sensing noise case, $G<0$ close to the transition for $N=547$, providing evidence for bistability in large enough systems.
For the actuation noise case, $N=547$ does not appear to be large enough to reach $G < 0$, but the dip in the transition region drops further and further below $G=1/3$ (the expected value for the disordered phase) as the system size is increased, suggesting that $G$ will reach negative values in larger systems \cite{BiRPP1997,GrChPRL2004,ChRaPRE2008}.
These numerical tests (and others we performed with different noise types and agent configurations) show that the AES transition is first order and has a bistable region.

An interesting aspect of the AES model is that we can use a continuous elastic sheet approximation to perform analytical calculations.
We follow this approach to carry out a standard linear stability analysis \cite{fetter2003theoretical} of the translating CM state in the zero noise case.
We begin by writing the elastic forces $\vec{F} = (F_x, F_y)$ that result from small displacements $\vec{u} = (u_x, u_y)$ of points on the membrane with respect to their equilibrium positions
\begin{eqnarray}
F_x &=& \left( \lambda + 2 \mu \right) \frac{\partial^2 u_x}{\partial x^2} + 
\mu \frac{\partial^2 u_x}{\partial y^2} + 
\left( \lambda + \mu \right) \frac{\partial^2 u_y}{\partial x \partial y}, 
\label{Eq:elasticityX} \\
F_y &=& \left( \lambda + 2 \mu \right) \frac{\partial^2 u_y}{\partial y^2} +
	\mu \frac{\partial^2 u_y}{\partial x^2} +
	\left( \lambda + \mu \right) \frac{\partial^2 u_x}{\partial x \partial y},
\label{Eq:elasticityY}
\end{eqnarray}
where the elastic constants are given by the Lam\'e parameter $\lambda$ and shear modulus $\mu$ \cite{fetter2003theoretical}.
We then linearize Eqs.~(\ref{Eq:dynX}) and (\ref{Eq:dynTheta}) around an equilibrium solution with undeformed membrane  and all agents moving at speed $v_0$ in the $\hat{x}$ direction, obtaining
$ \dot{u}_x = \alpha \, F_x$, $\dot{u}_y = v_0 \, \phi$, and $\dot{\phi} = \beta \, F_y$, 
where $\phi$ denotes perturbations to the $\theta = 0$ equilibrium heading angle.
Casting these expressions in Fourier space with wavevector components $(k_x,k_y)$, we can write the perturbation dynamics in matrix form and compute its eigenvalues $\Lambda$ to determine stability.
These are found to satisfy the characteristic polynomial equation 
$\Lambda^3 + C_2 \Lambda^2 + C_1 \Lambda + C_0 = 0$, with
\begin{eqnarray}
\label{Eq:charPolyCoeff0}
C_0 &=& \alpha \beta \mu v_0 \left( \lambda + 2 \mu \right) \left[ k_x^2 + k_y^2 \right]^2 , \\
\label{Eq:charPolyCoeff1}
C_1 &=& \beta v_0 \left[ \mu k_x^2 + \left( \lambda + 2 \mu \right) k_y^2 \right] , \\
\label{Eq:charPolyCoeff2}
C_2 &=& \alpha \left[ \left( \lambda + 2 \mu \right) k_x^2 + \mu k_y^2 \right].
\end{eqnarray}
Using Routh's stability criterion (here given by $C_1 C_2 > C_0$ \cite{Ogata}), we find that the system is stable if 
$\alpha \beta v_0 \left( \lambda + \mu \right)^2 k_x^2 k_y^2 > 0$, which is always verified.
We conclude that translating CM solutions are always linearly stable.
This is not the case, however, for most variations of the AES model.
For example, if we consider a constant speed algorithm by setting $\alpha = 0$, the characteristic polynomial becomes
$\Lambda^3 + \beta \, v_0 \left[ \mu k_x^2 + \left( \lambda + 2 \mu \right) k_y^2 \right] \Lambda = 0$,
which only has null or imaginary solutions.
Linear perturbations will therefore not dampen out, but produce instead permanent oscillations.
Numerical simulations confirm that, even for zero noise and starting from a perfectly aligned initial condition, the group loses order and agents start rotating in place instead of aligning.

\begin{figure}[t]
\includegraphics[width=86mm]{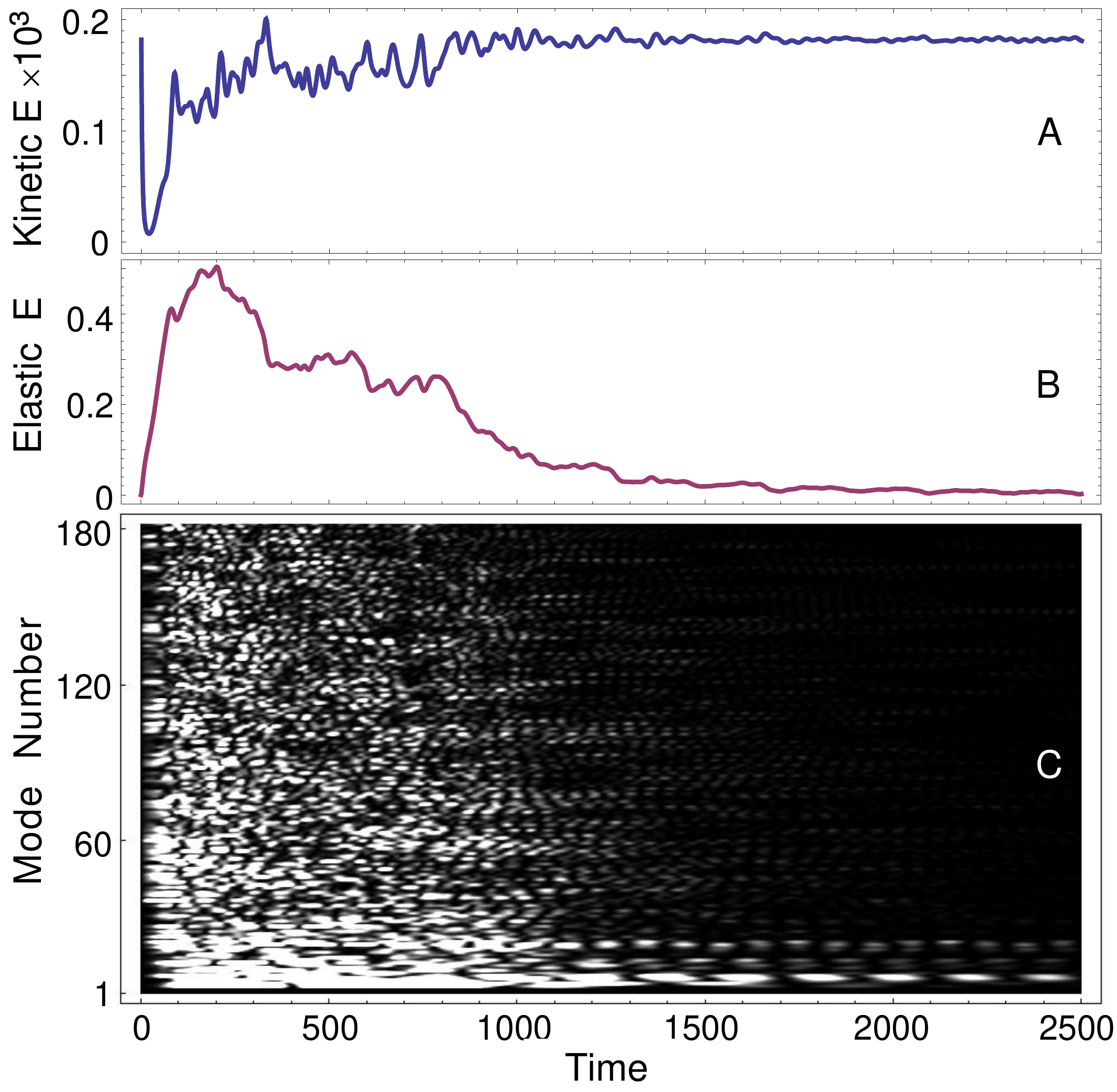}
\caption{\label{Fig3} (color online). 
Kinetic energy (A), elastic energy (B), and spectral decomposition of the elastic energy (C) as a function of time 
for an hexagonal $N=91$ active crystal simulation
(with zero noise and same initial condition as in panel A1 of Fig.~\ref{Fig1})
that converges to the aligned, translating state.
Brighter points on C indicate higher energies.
After an initial transient, A and B converge to their stationary values for collective translational motion.
All modes display energy levels that oscillate as they decay, with higher modes decaying faster.
Elastic energy flows to lower modes, producing coherent motion that eventually 
reaches the lowest (translational or rotational) modes.
}
\end{figure}
%

We now characterize the nonlinear energy cascading mechanism that drives the AES model to self-organize.
Figure \ref{Fig3} presents the energy dynamics of an hexagonal active crystal simulation with $N=91$ and zero noise
that converges to a translating solution.
Top panels display the total kinetic and potential energies as a function of time.
Panel C shows the spectral decomposition of the latter into its elastic normal modes, listed in order of 
growing energy and without accounting for degeneracies.
It is produced by first computing all $182$ elastic normal modes numerically (without considering agent orientations
or self-propulsion) and then expanding the dynamics into this basis.
The initial condition is set as in panel A1 of Fig.~\ref{Fig1}, with zero potential energy and 
kinetic energy $E_k = N v_0^2 / 2 = 1.82 \times 10^{-4}$ (setting the agent mass to $1$).
As the membrane deforms during the initial transient, potential energy grows and becomes broadly distributed over 
all modes (as expected for disordered systems), while kinetic energy drops. 
As time advances, the system rearranges itself into configurations with lower elastic energy and higher kinetic energy,
eventually reaching again (now in the ordered, translating state) values close to zero and $E_k$, respectively.
The energy of each mode oscillates while decaying, as in an underdamped oscillator, with higher modes decaying 
faster than lower ones.
This results from a combination of standard elasticity, self-propulsion, and the coupling between elastic forces and 
turning rate imposed by the AES model.
Indeed, in standard damped elastic systems, higher energy modes also decay faster than lower ones. 
Here, however, each agent is continuously injecting energy through its self-propulsion term, so motion cannot be fully dampened. 
Instead, modes decay by steering agents away from them.
Self-propulsion thus feeds energy to lower and lower modes, eventually reaching translational or rotational modes and achieving CM. 
Note that, despite this mechanism, similar models may not converge to CM. This will occur, for example, if agents inject too much energy into high-energy modes while turning (as in the $\alpha = 0$ case described above) or if they overshoot the angles that dampen high-energy modes by instantaneously switching heading (as in \cite{Gregoire2003157,GrChPRL2004}) instead of integrating Eq.~(\ref{Eq:dynTheta}).

We have identified in this Letter an alternative, elasticity-based mechanism that, in contrast to the Vicsek case, 
requires no exchange of heading information to achieve CM.
It also requires no switching of interacting neighbors over time to overcome the Mermin-Wagner theorem and achieve long-range order at non-zero noise levels \cite{MerminWagnerPRL1966,CzViJPhysA1997,ToTuPRL95,AlHuJSP2003}.
Instead, despite including no fluid-like mixing, we found that the AES model displays no loss of long-range order for larger structures, 
although the energy cascading mechanism takes longer to converge.
We found only one other model, introduced in \cite{SzSzPRE06} to study the collective migration of tissue cells, that can display CM under similar conditions.
In a version of this algorithm (designed to study active jamming) where agents only have repulsive interactions and are confined to a circular box, a similar elasticity-based mechanism was shown to be responsible for the dynamics of the jammed phase \cite{HeFiPRE11}.
While that model describes a different situation, where interaction forces can displace agents perpendicular to their heading, its detailed comparison to the AES model should yield a more complete understanding of the energy cascading mechanism.

Given that some kind of attraction-repulsion interaction must be present for any group of moving agents to remain cohesive and avoid overlapping, we expect the elasticity-based mechanism introduced here to play a relevant role in the CM dynamics of a variety of systems.
These could include microscopic biological or robotic agents (or even collectively migrating tissue cells \cite{SzSzPRE06,Trepat09physicalforces}) that can exert attraction-repulsion physical forces while being too simple to exhibit explicit aligning interactions.
We expect most animal groups to typically combine elasticity-based and alignment-based mechanisms in their CM dynamics, thus effectively behaving as an active viscoelastic material composed of aligning spins.
An interesting open question is the extent to which each mechanism is responsible for the CM of specific real-world swarms, 
which can depend on the time-scale and dynamical state considered.
Note that each mechanism results from interactions that produce different dynamical signatures, such as the properties of propagating waves or the response to perturbations.
Now that there is a growing number of experiments that allow the precise tracking of different types of swarms \cite{Cavagna2008217,KaHuCoPNAS2011}, 
these signatures could help determine their individual interaction rules based only on observed properties of their collective dynamics.

\begin{acknowledgments}
The work of CH was supported by the National Science Foundation under Grant No. PHY-0848755.
The work of EF and AT was partially supported by the Vlaanderen Research Foundation Flanders (Flemish Community of Belgium) through the H2Swarm project.
The work of EF, AT and MD was partially supported by the European Union's ERC Advanced Grant (contract 246939).
We also acknowledge support by the Max Planck Institute for the Physics of Complex Systems in Dresden, Germany, 
through the Advanced Study Group ``Statistical Physics of Collective Motion'', where part of this work was conducted.
\end{acknowledgments}

\bibliography{PRL_ActiveCrystals}

\begin{thebibliography}{31}%
\makeatletter
\providecommand \@ifxundefined [1]{%
 \@ifx{#1\undefined}
}%
\providecommand \@ifnum [1]{%
 \ifnum #1\expandafter \@firstoftwo
 \else \expandafter \@secondoftwo
 \fi
}%
\providecommand \@ifx [1]{%
 \ifx #1\expandafter \@firstoftwo
 \else \expandafter \@secondoftwo
 \fi
}%
\providecommand \natexlab [1]{#1}%
\providecommand \enquote  [1]{``#1''}%
\providecommand \bibnamefont  [1]{#1}%
\providecommand \bibfnamefont [1]{#1}%
\providecommand \citenamefont [1]{#1}%
\providecommand \href@noop [0]{\@secondoftwo}%
\providecommand \href [0]{\begingroup \@sanitize@url \@href}%
\providecommand \@href[1]{\@@startlink{#1}\@@href}%
\providecommand \@@href[1]{\endgroup#1\@@endlink}%
\providecommand \@sanitize@url [0]{\catcode `\\12\catcode `\$12\catcode
  `\&12\catcode `\#12\catcode `\^12\catcode `\_12\catcode `\%12\relax}%
\providecommand \@@startlink[1]{}%
\providecommand \@@endlink[0]{}%
\providecommand \url  [0]{\begingroup\@sanitize@url \@url }%
\providecommand \@url [1]{\endgroup\@href {#1}{\urlprefix }}%
\providecommand \urlprefix  [0]{URL }%
\providecommand \Eprint [0]{\href }%
\providecommand \doibase [0]{http://dx.doi.org/}%
\providecommand \selectlanguage [0]{\@gobble}%
\providecommand \bibinfo  [0]{\@secondoftwo}%
\providecommand \bibfield  [0]{\@secondoftwo}%
\providecommand \translation [1]{[#1]}%
\providecommand \BibitemOpen [0]{}%
\providecommand \bibitemStop [0]{}%
\providecommand \bibitemNoStop [0]{.\EOS\space}%
\providecommand \EOS [0]{\spacefactor3000\relax}%
\providecommand \BibitemShut  [1]{\csname bibitem#1\endcsname}%
\let\auto@bib@innerbib\@empty
\bibitem [{\citenamefont {Breder}(1954)}]{Breder54}%
  \BibitemOpen
  \bibfield  {author} {\bibinfo {author} {\bibfnamefont {C.}~\bibnamefont
  {Breder}},\ }\href {\doibase 10.2307/1930099} {\bibfield  {journal} {\bibinfo
   {journal} {Ecology}\ }\textbf {\bibinfo {volume} {35}},\ \bibinfo {pages}
  {361} (\bibinfo {year} {1954})}\BibitemShut {NoStop}%
\bibitem [{\citenamefont {Okubo}(1986)}]{Okubo86}%
  \BibitemOpen
  \bibfield  {author} {\bibinfo {author} {\bibfnamefont {A.}~\bibnamefont
  {Okubo}},\ }\href {\doibase 10.1016/0065-227X(86)90003-1} {\bibfield
  {journal} {\bibinfo  {journal} {Adv. Biophys.}\ }\textbf {\bibinfo {volume}
  {22}},\ \bibinfo {pages} {1} (\bibinfo {year} {1986})}\BibitemShut {NoStop}%
\bibitem [{\citenamefont {Gr\"unbaum}\ and\ \citenamefont
  {Okubo}(1994)}]{Grunbaum94}%
  \BibitemOpen
  \bibfield  {author} {\bibinfo {author} {\bibfnamefont {D.}~\bibnamefont
  {Gr\"unbaum}}\ and\ \bibinfo {author} {\bibfnamefont {A.}~\bibnamefont
  {Okubo}},\ }\bibfield  {booktitle} {\emph {\bibinfo {booktitle} {Frontiers in
  Theoretical Biology, Lecture notes in Biomathematics}},\ }\href@noop {} {\
  \textbf {\bibinfo {volume} {100}},\ \bibinfo {pages} {296} (\bibinfo {year}
  {1994})}\BibitemShut {NoStop}%
\bibitem [{\citenamefont {Shapiro}\ and\ \citenamefont
  {Dworkin}(1997)}]{ShapiroBook97}%
  \BibitemOpen
  \bibinfo {editor} {\bibfnamefont {J.~A.}\ \bibnamefont {Shapiro}}\ and\
  \bibinfo {editor} {\bibfnamefont {M.}~\bibnamefont {Dworkin}},\ eds.,\
  \href@noop {} {\emph {\bibinfo {title} {{Bacteria as multicellular
  organisms}}}}\ (\bibinfo  {publisher} {{Oxford University Press}},\ \bibinfo
  {year} {1997})\BibitemShut {NoStop}%
\bibitem [{\citenamefont {Couzin}\ \emph {et~al.}(2002)\citenamefont {Couzin},
  \citenamefont {Krause}, \citenamefont {James}, \citenamefont {Ruxton},\ and\
  \citenamefont {Franks}}]{Couzin2002}%
  \BibitemOpen
  \bibfield  {author} {\bibinfo {author} {\bibfnamefont {I.~D.}\ \bibnamefont
  {Couzin}}, \bibinfo {author} {\bibfnamefont {J.}~\bibnamefont {Krause}},
  \bibinfo {author} {\bibfnamefont {R.}~\bibnamefont {James}}, \bibinfo
  {author} {\bibfnamefont {G.~D.}\ \bibnamefont {Ruxton}}, \ and\ \bibinfo
  {author} {\bibfnamefont {N.~R.}\ \bibnamefont {Franks}},\ }\href {\doibase
  10.1006/jtbi.2002.3065} {\bibfield  {journal} {\bibinfo  {journal} {J. Theor.
  Biol.}\ }\textbf {\bibinfo {volume} {218}},\ \bibinfo {pages} {1 } (\bibinfo
  {year} {2002})}\BibitemShut {NoStop}%
\bibitem [{\citenamefont {Sumpter}(2010)}]{SumpterBook}%
  \BibitemOpen
  \bibfield  {author} {\bibinfo {author} {\bibfnamefont {D.~J.~T.}\
  \bibnamefont {Sumpter}},\ }\href@noop {} {\emph {\bibinfo {title}
  {{Collective Animal Behavior}}}}\ (\bibinfo  {publisher} {{Princeton
  University Press}},\ \bibinfo {year} {2010})\BibitemShut {NoStop}%
\bibitem [{\citenamefont {Deutsch}\ \emph {et~al.}(2012)\citenamefont
  {Deutsch}, \citenamefont {Theraulaz},\ and\ \citenamefont
  {Vicsek}}]{DeThViReview2012}%
  \BibitemOpen
  \bibfield  {author} {\bibinfo {author} {\bibfnamefont {A.}~\bibnamefont
  {Deutsch}}, \bibinfo {author} {\bibfnamefont {G.}~\bibnamefont {Theraulaz}},
  \ and\ \bibinfo {author} {\bibfnamefont {T.}~\bibnamefont {Vicsek}},\ }\href
  {\doibase 10.1098/​rsfs.2012.0048} {\bibfield  {journal} {\bibinfo
  {journal} {Interface Focus}\ }\textbf {\bibinfo {volume} {2}},\ \bibinfo
  {pages} {689} (\bibinfo {year} {2012})}\BibitemShut {NoStop}%
\bibitem [{\citenamefont {Vicsek}\ and\ \citenamefont
  {Zafeiris}(2012)}]{VicsekZafeirisReview2012}%
  \BibitemOpen
  \bibfield  {author} {\bibinfo {author} {\bibfnamefont {T.}~\bibnamefont
  {Vicsek}}\ and\ \bibinfo {author} {\bibfnamefont {A.}~\bibnamefont
  {Zafeiris}},\ }\href {\doibase 10.1016/j.physrep.2012.03.004} {\bibfield
  {journal} {\bibinfo  {journal} {Phys. Rep.}\ }\textbf {\bibinfo {volume}
  {517}},\ \bibinfo {pages} {71 } (\bibinfo {year} {2012})}\BibitemShut
  {NoStop}%
\bibitem [{\citenamefont {Brambilla}\ \emph {et~al.}(tion)\citenamefont
  {Brambilla}, \citenamefont {Ferrante}, \citenamefont {Birattari},\ and\
  \citenamefont {Dorigo}}]{SwarmRoboticsReview2013}%
  \BibitemOpen
  \bibfield  {author} {\bibinfo {author} {\bibfnamefont {M.}~\bibnamefont
  {Brambilla}}, \bibinfo {author} {\bibfnamefont {E.}~\bibnamefont {Ferrante}},
  \bibinfo {author} {\bibfnamefont {M.}~\bibnamefont {Birattari}}, \ and\
  \bibinfo {author} {\bibfnamefont {M.}~\bibnamefont {Dorigo}},\ }\href@noop {}
  {\bibfield  {journal} {\bibinfo  {journal} {Swarm Intelligence}\ } (\bibinfo
  {year} {accepted for publication})}\BibitemShut {NoStop}%
\bibitem [{\citenamefont {Vicsek}\ \emph {et~al.}(1995)\citenamefont {Vicsek},
  \citenamefont {Czir\'ok}, \citenamefont {Ben-Jacob}, \citenamefont {Cohen},\
  and\ \citenamefont {Shochet}}]{ViCzPRL95}%
  \BibitemOpen
  \bibfield  {author} {\bibinfo {author} {\bibfnamefont {T.}~\bibnamefont
  {Vicsek}}, \bibinfo {author} {\bibfnamefont {A.}~\bibnamefont {Czir\'ok}},
  \bibinfo {author} {\bibfnamefont {E.}~\bibnamefont {Ben-Jacob}}, \bibinfo
  {author} {\bibfnamefont {I.}~\bibnamefont {Cohen}}, \ and\ \bibinfo {author}
  {\bibfnamefont {O.}~\bibnamefont {Shochet}},\ }\href {\doibase
  10.1103/PhysRevLett.75.1226} {\bibfield  {journal} {\bibinfo  {journal}
  {Phys. Rev. Lett.}\ }\textbf {\bibinfo {volume} {75}},\ \bibinfo {pages}
  {1226} (\bibinfo {year} {1995})}\BibitemShut {NoStop}%
\bibitem [{\citenamefont {Czir\'ok}\ \emph {et~al.}(1997)\citenamefont
  {Czir\'ok}, \citenamefont {Stanley},\ and\ \citenamefont
  {Vicsek}}]{CzViJPhysA1997}%
  \BibitemOpen
  \bibfield  {author} {\bibinfo {author} {\bibfnamefont {A.}~\bibnamefont
  {Czir\'ok}}, \bibinfo {author} {\bibfnamefont {H.~E.}\ \bibnamefont
  {Stanley}}, \ and\ \bibinfo {author} {\bibfnamefont {T.}~\bibnamefont
  {Vicsek}},\ }\href {http://stacks.iop.org/0305-4470/30/i=5/a=009} {\bibfield
  {journal} {\bibinfo  {journal} {J. Phys. A: Math. Gen.}\ }\textbf {\bibinfo
  {volume} {30}},\ \bibinfo {pages} {1375} (\bibinfo {year}
  {1997})}\BibitemShut {NoStop}%
\bibitem [{\citenamefont {Chat\'e}\ \emph {et~al.}(2008)\citenamefont
  {Chat\'e}, \citenamefont {Ginelli}, \citenamefont {Gr\'egoire},\ and\
  \citenamefont {Raynaud}}]{ChRaPRE2008}%
  \BibitemOpen
  \bibfield  {author} {\bibinfo {author} {\bibfnamefont {H.}~\bibnamefont
  {Chat\'e}}, \bibinfo {author} {\bibfnamefont {F.}~\bibnamefont {Ginelli}},
  \bibinfo {author} {\bibfnamefont {G.}~\bibnamefont {Gr\'egoire}}, \ and\
  \bibinfo {author} {\bibfnamefont {F.}~\bibnamefont {Raynaud}},\ }\href
  {\doibase 10.1103/PhysRevE.77.046113} {\bibfield  {journal} {\bibinfo
  {journal} {Phys. Rev. E}\ }\textbf {\bibinfo {volume} {77}},\ \bibinfo
  {pages} {046113} (\bibinfo {year} {2008})}\BibitemShut {NoStop}%
\bibitem [{\citenamefont {Binney}\ \emph {et~al.}(1992)\citenamefont {Binney},
  \citenamefont {Dowrick}, \citenamefont {Fisher},\ and\ \citenamefont
  {Newman}}]{CriticalPhenomenaBook}%
  \BibitemOpen
  \bibfield  {author} {\bibinfo {author} {\bibfnamefont {J.~J.}\ \bibnamefont
  {Binney}}, \bibinfo {author} {\bibfnamefont {N.~J.}\ \bibnamefont {Dowrick}},
  \bibinfo {author} {\bibfnamefont {A.~J.}\ \bibnamefont {Fisher}}, \ and\
  \bibinfo {author} {\bibfnamefont {M.~E.~J.}\ \bibnamefont {Newman}},\
  }\href@noop {} {\emph {\bibinfo {title} {{The Theory of Critical Phenomena:
  An Introduction to Renormalization Group}}}}\ (\bibinfo  {publisher} {{Oxford
  University Press}},\ \bibinfo {year} {1992})\BibitemShut {NoStop}%
\bibitem [{\citenamefont {Toner}\ and\ \citenamefont {Tu}(1995)}]{ToTuPRL95}%
  \BibitemOpen
  \bibfield  {author} {\bibinfo {author} {\bibfnamefont {J.}~\bibnamefont
  {Toner}}\ and\ \bibinfo {author} {\bibfnamefont {Y.}~\bibnamefont {Tu}},\
  }\href {\doibase 10.1103/PhysRevLett.75.4326} {\bibfield  {journal} {\bibinfo
   {journal} {Phys. Rev. Lett.}\ }\textbf {\bibinfo {volume} {75}},\ \bibinfo
  {pages} {4326} (\bibinfo {year} {1995})}\BibitemShut {NoStop}%
\bibitem [{\citenamefont {Toner}\ and\ \citenamefont {Tu}(1998)}]{ToTuPRE98}%
  \BibitemOpen
  \bibfield  {author} {\bibinfo {author} {\bibfnamefont {J.}~\bibnamefont
  {Toner}}\ and\ \bibinfo {author} {\bibfnamefont {Y.}~\bibnamefont {Tu}},\
  }\href {\doibase 10.1103/PhysRevE.58.4828} {\bibfield  {journal} {\bibinfo
  {journal} {Phys. Rev. E}\ }\textbf {\bibinfo {volume} {58}},\ \bibinfo
  {pages} {4828} (\bibinfo {year} {1998})}\BibitemShut {NoStop}%
\bibitem [{\citenamefont {Bertin}\ \emph {et~al.}(2006)\citenamefont {Bertin},
  \citenamefont {Droz},\ and\ \citenamefont {Gr\'egoire}}]{PhysRevE.74.022101}%
  \BibitemOpen
  \bibfield  {author} {\bibinfo {author} {\bibfnamefont {E.}~\bibnamefont
  {Bertin}}, \bibinfo {author} {\bibfnamefont {M.}~\bibnamefont {Droz}}, \ and\
  \bibinfo {author} {\bibfnamefont {G.}~\bibnamefont {Gr\'egoire}},\ }\href
  {\doibase 10.1103/PhysRevE.74.022101} {\bibfield  {journal} {\bibinfo
  {journal} {Phys. Rev. E}\ }\textbf {\bibinfo {volume} {74}},\ \bibinfo
  {pages} {022101} (\bibinfo {year} {2006})}\BibitemShut {NoStop}%
\bibitem [{\citenamefont {Peshkov}\ \emph {et~al.}(2012)\citenamefont
  {Peshkov}, \citenamefont {Ngo}, \citenamefont {Bertin}, \citenamefont
  {Chat\'e},\ and\ \citenamefont {Ginelli}}]{PhysRevLett.109.098101}%
  \BibitemOpen
  \bibfield  {author} {\bibinfo {author} {\bibfnamefont {A.}~\bibnamefont
  {Peshkov}}, \bibinfo {author} {\bibfnamefont {S.}~\bibnamefont {Ngo}},
  \bibinfo {author} {\bibfnamefont {E.}~\bibnamefont {Bertin}}, \bibinfo
  {author} {\bibfnamefont {H.}~\bibnamefont {Chat\'e}}, \ and\ \bibinfo
  {author} {\bibfnamefont {F.}~\bibnamefont {Ginelli}},\ }\href {\doibase
  10.1103/PhysRevLett.109.098101} {\bibfield  {journal} {\bibinfo  {journal}
  {Phys. Rev. Lett.}\ }\textbf {\bibinfo {volume} {109}},\ \bibinfo {pages}
  {098101} (\bibinfo {year} {2012})}\BibitemShut {NoStop}%
\bibitem [{\citenamefont {Romanczuk}\ \emph {et~al.}(2009)\citenamefont
  {Romanczuk}, \citenamefont {Couzin},\ and\ \citenamefont
  {Schimansky-Geier}}]{RoCoPRL09}%
  \BibitemOpen
  \bibfield  {author} {\bibinfo {author} {\bibfnamefont {P.}~\bibnamefont
  {Romanczuk}}, \bibinfo {author} {\bibfnamefont {I.~D.}\ \bibnamefont
  {Couzin}}, \ and\ \bibinfo {author} {\bibfnamefont {L.}~\bibnamefont
  {Schimansky-Geier}},\ }\href {\doibase 10.1103/PhysRevLett.102.010602}
  {\bibfield  {journal} {\bibinfo  {journal} {Phys. Rev. Lett.}\ }\textbf
  {\bibinfo {volume} {102}},\ \bibinfo {pages} {010602} (\bibinfo {year}
  {2009})}\BibitemShut {NoStop}%
\bibitem [{\citenamefont {Grossman}\ \emph {et~al.}(2008)\citenamefont
  {Grossman}, \citenamefont {Aranson},\ and\ \citenamefont
  {Jacob}}]{GrArNJP08}%
  \BibitemOpen
  \bibfield  {author} {\bibinfo {author} {\bibfnamefont {D.}~\bibnamefont
  {Grossman}}, \bibinfo {author} {\bibfnamefont {I.~S.}\ \bibnamefont
  {Aranson}}, \ and\ \bibinfo {author} {\bibfnamefont {E.~B.}\ \bibnamefont
  {Jacob}},\ }\href@noop {} {\bibfield  {journal} {\bibinfo  {journal} {New J.
  Phys.}\ }\textbf {\bibinfo {volume} {10}},\ \bibinfo {pages} {023036}
  (\bibinfo {year} {2008})}\BibitemShut {NoStop}%
\bibitem [{\citenamefont {Szab\'o}\ \emph {et~al.}(2006)\citenamefont
  {Szab\'o}, \citenamefont {Sz\"oll\"osi}, \citenamefont {G\"onci},
  \citenamefont {Jur\'anyi}, \citenamefont {Selmeczi},\ and\ \citenamefont
  {Vicsek}}]{SzSzPRE06}%
  \BibitemOpen
  \bibfield  {author} {\bibinfo {author} {\bibfnamefont {B.}~\bibnamefont
  {Szab\'o}}, \bibinfo {author} {\bibfnamefont {G.~J.}\ \bibnamefont
  {Sz\"oll\"osi}}, \bibinfo {author} {\bibfnamefont {B.}~\bibnamefont
  {G\"onci}}, \bibinfo {author} {\bibfnamefont {Z.}~\bibnamefont {Jur\'anyi}},
  \bibinfo {author} {\bibfnamefont {D.}~\bibnamefont {Selmeczi}}, \ and\
  \bibinfo {author} {\bibfnamefont {T.}~\bibnamefont {Vicsek}},\ }\href@noop {}
  {\bibfield  {journal} {\bibinfo  {journal} {Phys. Rev. E}\ }\textbf {\bibinfo
  {volume} {74}},\ \bibinfo {pages} {061908} (\bibinfo {year}
  {2006})}\BibitemShut {NoStop}%
\bibitem [{\citenamefont {Henkes}\ \emph {et~al.}(2011)\citenamefont {Henkes},
  \citenamefont {Fily},\ and\ \citenamefont {Marchetti}}]{HeFiPRE11}%
  \BibitemOpen
  \bibfield  {author} {\bibinfo {author} {\bibfnamefont {S.}~\bibnamefont
  {Henkes}}, \bibinfo {author} {\bibfnamefont {Y.}~\bibnamefont {Fily}}, \ and\
  \bibinfo {author} {\bibfnamefont {M.~C.}\ \bibnamefont {Marchetti}},\
  }\href@noop {} {\bibfield  {journal} {\bibinfo  {journal} {Phys. Rev. E}\
  }\textbf {\bibinfo {volume} {84}},\ \bibinfo {pages} {040301(R)} (\bibinfo
  {year} {2011})}\BibitemShut {NoStop}%
\bibitem [{\citenamefont {Gr\'egoire}\ \emph {et~al.}(2003)\citenamefont
  {Gr\'egoire}, \citenamefont {Chat\'e},\ and\ \citenamefont
  {Tu}}]{Gregoire2003157}%
  \BibitemOpen
  \bibfield  {author} {\bibinfo {author} {\bibfnamefont {G.}~\bibnamefont
  {Gr\'egoire}}, \bibinfo {author} {\bibfnamefont {H.}~\bibnamefont {Chat\'e}},
  \ and\ \bibinfo {author} {\bibfnamefont {Y.}~\bibnamefont {Tu}},\ }\href
  {\doibase 10.1016/S0167-2789(03)00102-7} {\bibfield  {journal} {\bibinfo
  {journal} {Physica D}\ }\textbf {\bibinfo {volume} {181}},\ \bibinfo {pages}
  {157 } (\bibinfo {year} {2003})}\BibitemShut {NoStop}%
\bibitem [{\citenamefont {Gr\'egoire}\ and\ \citenamefont
  {Chat\'e}(2004)}]{GrChPRL2004}%
  \BibitemOpen
  \bibfield  {author} {\bibinfo {author} {\bibfnamefont {G.}~\bibnamefont
  {Gr\'egoire}}\ and\ \bibinfo {author} {\bibfnamefont {H.}~\bibnamefont
  {Chat\'e}},\ }\href {\doibase 10.1103/PhysRevLett.92.025702} {\bibfield
  {journal} {\bibinfo  {journal} {Phys. Rev. Lett.}\ }\textbf {\bibinfo
  {volume} {92}},\ \bibinfo {pages} {025702} (\bibinfo {year}
  {2004})}\BibitemShut {NoStop}%
\bibitem [{\citenamefont {Fetter}\ and\ \citenamefont
  {Walecka}(2003)}]{fetter2003theoretical}%
  \BibitemOpen
  \bibfield  {author} {\bibinfo {author} {\bibfnamefont {A.}~\bibnamefont
  {Fetter}}\ and\ \bibinfo {author} {\bibfnamefont {J.}~\bibnamefont
  {Walecka}},\ }\href {http://books.google.com.tr/books?id=olMpStYOlnoC} {\emph
  {\bibinfo {title} {Theoretical Mechanics of Particles and Continua}}},\ Dover
  Books on Physics Series\ (\bibinfo  {publisher} {Dover},\ \bibinfo {year}
  {2003})\BibitemShut {NoStop}%
\bibitem [{\citenamefont {Aldana}\ and\ \citenamefont
  {Huepe}(2003)}]{AlHuJSP2003}%
  \BibitemOpen
  \bibfield  {author} {\bibinfo {author} {\bibfnamefont {M.}~\bibnamefont
  {Aldana}}\ and\ \bibinfo {author} {\bibfnamefont {C.}~\bibnamefont {Huepe}},\
  }\href@noop {} {\bibfield  {journal} {\bibinfo  {journal} {J. Stat. Phys.}\
  }\textbf {\bibinfo {volume} {112}},\ \bibinfo {pages} {135} (\bibinfo {year}
  {2003})}\BibitemShut {NoStop}%
\bibitem [{\citenamefont {Binder}(1997)}]{BiRPP1997}%
  \BibitemOpen
  \bibfield  {author} {\bibinfo {author} {\bibfnamefont {K.}~\bibnamefont
  {Binder}},\ }\href {http://stacks.iop.org/0034-4885/60/i=5/a=001} {\bibfield
  {journal} {\bibinfo  {journal} {Rep. Prog. Phys.}\ }\textbf {\bibinfo
  {volume} {60}},\ \bibinfo {pages} {487} (\bibinfo {year} {1997})}\BibitemShut
  {NoStop}%
\bibitem [{\citenamefont {Ogata}(2001)}]{Ogata}%
  \BibitemOpen
  \bibfield  {author} {\bibinfo {author} {\bibfnamefont {K.}~\bibnamefont
  {Ogata}},\ }\href@noop {} {\emph {\bibinfo {title} {{Modern Control
  Engineering (4th Edition)}}}}\ (\bibinfo  {publisher} {{Prentice Hall}},\
  \bibinfo {year} {2001})\BibitemShut {NoStop}%
\bibitem [{\citenamefont {Mermin}\ and\ \citenamefont
  {Wagner}(1966)}]{MerminWagnerPRL1966}%
  \BibitemOpen
  \bibfield  {author} {\bibinfo {author} {\bibfnamefont {N.~D.}\ \bibnamefont
  {Mermin}}\ and\ \bibinfo {author} {\bibfnamefont {H.}~\bibnamefont
  {Wagner}},\ }\href {\doibase 10.1103/PhysRevLett.17.1133} {\bibfield
  {journal} {\bibinfo  {journal} {Phys. Rev. Lett.}\ }\textbf {\bibinfo
  {volume} {17}},\ \bibinfo {pages} {1133} (\bibinfo {year}
  {1966})}\BibitemShut {NoStop}%
\bibitem [{\citenamefont {Trepat}\ \emph {et~al.}(2009)\citenamefont {Trepat},
  \citenamefont {Wasserman}, \citenamefont {Angelini}, \citenamefont {Millet},
  \citenamefont {Weitz}, \citenamefont {Butler},\ and\ \citenamefont
  {Fredberg}}]{Trepat09physicalforces}%
  \BibitemOpen
  \bibfield  {author} {\bibinfo {author} {\bibfnamefont {X.}~\bibnamefont
  {Trepat}}, \bibinfo {author} {\bibfnamefont {M.~R.}\ \bibnamefont
  {Wasserman}}, \bibinfo {author} {\bibfnamefont {T.~E.}\ \bibnamefont
  {Angelini}}, \bibinfo {author} {\bibfnamefont {E.}~\bibnamefont {Millet}},
  \bibinfo {author} {\bibfnamefont {D.~A.}\ \bibnamefont {Weitz}}, \bibinfo
  {author} {\bibfnamefont {J.~P.}\ \bibnamefont {Butler}}, \ and\ \bibinfo
  {author} {\bibfnamefont {J.~J.}\ \bibnamefont {Fredberg}},\ }\href {\doibase
  10.1038/nphys1269} {\bibfield  {journal} {\bibinfo  {journal} {Nat. Phys.}\
  }\textbf {\bibinfo {volume} {5}},\ \bibinfo {pages} {426} (\bibinfo {year}
  {2009})}\BibitemShut {NoStop}%
\bibitem [{\citenamefont {Cavagna}\ \emph {et~al.}(2008)\citenamefont
  {Cavagna}, \citenamefont {Giardina}, \citenamefont {Orlandi}, \citenamefont
  {Parisi}, \citenamefont {Procaccini}, \citenamefont {Viale},\ and\
  \citenamefont {Zdravkovic}}]{Cavagna2008217}%
  \BibitemOpen
  \bibfield  {author} {\bibinfo {author} {\bibfnamefont {A.}~\bibnamefont
  {Cavagna}}, \bibinfo {author} {\bibfnamefont {I.}~\bibnamefont {Giardina}},
  \bibinfo {author} {\bibfnamefont {A.}~\bibnamefont {Orlandi}}, \bibinfo
  {author} {\bibfnamefont {G.}~\bibnamefont {Parisi}}, \bibinfo {author}
  {\bibfnamefont {A.}~\bibnamefont {Procaccini}}, \bibinfo {author}
  {\bibfnamefont {M.}~\bibnamefont {Viale}}, \ and\ \bibinfo {author}
  {\bibfnamefont {V.}~\bibnamefont {Zdravkovic}},\ }\href {\doibase
  10.1016/j.anbehav.2008.02.002} {\bibfield  {journal} {\bibinfo  {journal}
  {Animal Behaviour}\ }\textbf {\bibinfo {volume} {76}},\ \bibinfo {pages} {217
  } (\bibinfo {year} {2008})}\BibitemShut {NoStop}%
\bibitem [{\citenamefont {Katz}\ \emph {et~al.}(2011)\citenamefont {Katz},
  \citenamefont {Tunstrom}, \citenamefont {Ioannou}, \citenamefont {Huepe},\
  and\ \citenamefont {Couzin}}]{KaHuCoPNAS2011}%
  \BibitemOpen
  \bibfield  {author} {\bibinfo {author} {\bibfnamefont {Y.}~\bibnamefont
  {Katz}}, \bibinfo {author} {\bibfnamefont {K.}~\bibnamefont {Tunstrom}},
  \bibinfo {author} {\bibfnamefont {C.~C.}\ \bibnamefont {Ioannou}}, \bibinfo
  {author} {\bibfnamefont {C.}~\bibnamefont {Huepe}}, \ and\ \bibinfo {author}
  {\bibfnamefont {I.~D.}\ \bibnamefont {Couzin}},\ }\href@noop {} {\bibfield
  {journal} {\bibinfo  {journal} {P. Natl. Acad. Sci. USA}\ }\textbf {\bibinfo
  {volume} {108}},\ \bibinfo {pages} {18720} (\bibinfo {year}
  {2011})}\BibitemShut {NoStop}%
\end{thebibliography}%

\end{document}